\documentclass[USenglish,oneside,twocolumn]{article}

\usepackage[utf8]{inputenc}
\usepackage[big]{dgruyter_NEW}

\usepackage{amsfonts}
\usepackage{amsmath,amssymb}
\usepackage{multirow}
\usepackage{url}
\usepackage{ulem}
\usepackage{algorithm}
\usepackage{algorithmic}
\usepackage{color}
\usepackage{xcolor}
 
\DOI{foobar}

\cclogo{\includegraphics{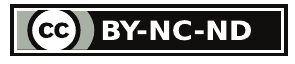}}

\newcommand{\argmax}{\mathop{\rm argmax}\limits}

\makeatletter
\def\plist@algorithm{Alg.\space}
\makeatother

\makeatletter
\def\blfootnote{\xdef\@thefnmark{}\@footnotetext}
\makeatother

\begin{document}
  \author[1]{Rishav Chourasia}
  \author[2]{Batnyam Enkhtaivan}
  \author[3]{Kunihiro Ito}
  \author[4]{Junki Mori}
  \author[5]{Isamu Teranishi}
  \author[6]{Hikaru Tsuchida}
  \affil[1]{National University of Singapore, E-mail: \url{e0427764@u.nus.edu}}

  \affil[2]{NEC Corporation, E-mail: \url{b-enkhtaivan@nec.com}}

  \affil[3]{NEC Corporation, E-mail: \url{kunihiro.ito.220@nec.com}}

  \affil[4]{NEC Corporation, E-mail: \url{junki.mori@nec.com}}

  \affil[5]{NEC Corporation, E-mail: \url{teranisi@nec.com}}

  \affil[6]{NEC Corporation, E-mail: \url{h_tsuchida@nec.com}}

  \title{\huge Knowledge Cross-Distillation for Membership Privacy}

  \runningtitle{Knowledge Cross-Distillation for Membership Privacy}



\begin{abstract}
{A \textit{membership inference attack (MIA)} poses privacy risks for the training data of a machine learning model. With an MIA, an attacker guesses if the target data are a member of the training dataset. The state-of-the-art defense against MIAs, distillation for membership privacy (DMP), requires not only private data for protection but a large amount of unlabeled public data. However, in certain privacy-sensitive domains, such as medicine and finance, the availability of public data is not guaranteed. Moreover, a trivial method for generating public data by using generative adversarial networks significantly decreases the model accuracy, as reported by the authors of DMP. To overcome this problem, we propose a novel defense against MIAs that uses knowledge distillation without requiring public data. Our experiments show that the privacy protection and accuracy of our defense are comparable to those of DMP for the benchmark tabular datasets used in MIA research, Purchase100 and Texas100, and our defense has a much better privacy-utility trade-off than those of the existing defenses that also do not use  public data for the image dataset CIFAR10.}
\end{abstract}


  \journalname{Proceedings on Privacy Enhancing Technologies}
\DOI{Editor to enter DOI}
  \startpage{1}
  \received{..}
  \revised{..}
  \accepted{..}

  \journalyear{..}
  \journalvolume{..}
  \journalissue{..}

\maketitle

\section{Introduction} 
\label{sec:introduction}
\subsection{Background}
Machine learning (ML) has been extensively used in various aspects of society \cite{dlsurvey}. We have seen great improvements in areas such as image recognition and natural language processing.
\blfootnote{The authors are alphabetically ordered.}

However, in the recent years, it has been reported that the privacy of the training data can be significantly undermined by analyzing ML models. Since, in most applications, privacy-sensitive data are used as the training data for the models, protecting the privacy of the training data is crucial for getting approval from data providers or essentially society.

Following the growing concern for privacy in society worldwide, many countries and regions are introducing regulations for data protection, e.g., the General Data Protection Regulation (GDPR) \cite{GDPR}, California Consumer Privacy Act (CCPA) \cite{CCPA}, and Health Insurance Portability and Accountability Act (HIPAA) \cite{HIPAA}. Moreover, guidelines and regulations designed specifically for trustworthiness in artificial intelligence (AI) and ML are under discussion \cite{EU-Guideline}.

\vspace{1em}\noindent\textbf{Membership Inference Attacks:} One of the most fundamental attacks against the privacy of a ML model is the \textit{membership inference attack (MIA)} \cite{shokri2017membership,nasr2018machine,salem2019ml,song2020systematic,DBLP:conf/sp/NasrSH19,DBLP:conf/ccs/SongSM19,choo2020label,veale2018algorithms,yeom2018privacy,DBLP:journals/corr/abs-2103-07853,sablayrolles2019white}, where an attacker guesses whether the given target data is in the training data of a ML model. 

MIAs are dangerous because they reveal the information of individual pieces of data rather than the trend of the whole population of training data. For instance, consider an ML model for inferring a reaction to some drug from a cancer patient's morphological data. An MIA attacker who knows the victim's data and has access rights to the ML model can know whether the victim has cancer or not, although the victim's data itself do not directly contain this information.

Another reason that MIAs are dangerous is that they can be executed through legitimate access to ML models only, meaning that they cannot be prevented by the conventional security methods such as data encryption and access control \cite{shokri2017membership}.

\vspace{1em}\noindent\textbf{Defense against MIAs:} The current state-of-the-art defense against MIAs is \textit{Distillation for Membership Privacy (DMP)} \cite{shejwalkar2021membership}. It can protect even against various state-of-the-art MIA attacks \cite{song2020systematic,DBLP:conf/sp/NasrSH19,choo2020label}, which the previous defenses \cite{nasr2018machine,jia2019memguard} cannot protect against very well, and its success comes from the ``semi-supervised assumption'' that a defender can obtain public unlabeled data. Specifically, DMP exploits a knowledge transfer technique \cite{hinton2015distil}; a defender trains an ML model using their own private data, feeds public data to the ML model to obtain the outputs of them, and trains another ML model using the public data and the corresponding outputs. Such indirect usage of private data makes knowledge distillation-based methods highly effective in protecting the privacy of private data.

However, in many domains of ML applications, public data are scarce due to the sensitive nature of the data, e.g., financial and medical data. To overcome this, utilization of synthetic data is proposed \cite{shejwalkar2021membership} as well. However, this method decreases accuracy \cite{shejwalkar2021membership} due to the decrease in data quality.

\subsection{Our Contributions}
In this paper, we propose a novel knowledge distillation-based defense that uses only private data for model training. 

Our contributions are as follows.
\begin{itemize}
\item We propose a novel MIA defense called \textit{knowledge cross-distillation (KCD)}\footnote{After we submitted our work to PETS 2022 Issue 2, Tang et al. \cite{tang2021mitigating} published a concurrent and independent work similar to ours in arXiv.}. Unlike the state-of-the-art defense, DMP, it does not require any public or synthetic reference data to protect ML models. Hence, KCD allows us to protect the privacy of ML models in areas where public reference data are scarce.
\item For the benchmark tabular datasets used in MIA research, Purchase100 and Texas100, we empirically show that the privacy protection and accuracy of KCD are comparable to those of DMP even though KCD does not require public or synthetic data, unlike DMP.
\item For the image dataset CIFAR10, we empirically show that the accuracy of KCD is comparable to that of DMP, and KCD provides a much better privacy-utility trade-off than those of other defenses that do not require public or synthetic reference data.
\end{itemize}

\subsection{Other Related Works}
We focus only on related works that are directly related to our contributions. See Hu et al. \cite{DBLP:journals/corr/abs-2103-07853} for a comprehensive survey of MIAs.

\vspace{1em}
\noindent\textbf{Membership Inference Attacks:} One of the earliest works considering MIAs is by Homer et al. \cite{homer2008resolving}, and MIAs were introduced in the ML setting in a seminal work by Shokri et al. \cite{shokri2017membership}. A series of MIA attacks, which is now called the \textit{neural network-based attack}, was proposed by Shokri et al. \cite{shokri2017membership} and was studied in detail by Salem et al. \cite{salem2019ml} and Truex et al. \cite{8634878}. Later, a new type of MIA attack, the \textit{metric-based attack}, was proposed by Yeom et al. \cite{yeom2018privacy} and studied by Song et al. \cite{DBLP:conf/ccs/SongSM19}, Salem et al. \cite{salem2019ml}, and Leino et al. \cite{DBLP:conf/uss/LeinoF20}. Then, Song et al. \cite{song2020systematic} summarized and improved upon them and proposed the state-of-the art metric-based attack as well.
 
Choo et al. \cite{choo2020label} and Li et al. \cite{DBLP:journals/corr/abs-2007-15528} independently and concurrently succeeded in attacking neural networks in a \textit{label-only setting}, where an attacker can get only labels as outputs of a target neural network, while the attackers of other known papers require confidence scores as the outputs of it. Nasr et al. \cite{DBLP:conf/sp/NasrSH19} proposed an MIA attack in a \textit{white-box setting}, where an attacker can obtain the structure and parameters of the target neural network.

\vspace{1em}

\noindent\textbf{Known Defenses:} MIAs can be mitigated using one known method, \textit{differential privacy} \cite{10.1007/11787006_1,DBLP:conf/eurocrypt/DworkKMMN06}, which is a technique for guaranteeing worst-case privacy by adding noise to the learning objective or model outputs.
However, defenses designed to protect against MIAs specifically have better privacy-utility trade-offs. Three MIA-specific defenses were proposed: AdvReg by Nasr et al. \cite{nasr2018machine}, MemGuard by Jia et al. \cite{jia2019memguard}, and DMP \cite{shejwalkar2021membership}.

An important technique for protecting MLs against MIAs is knowledge transfer \cite{hinton2015distil}. Using this technique, PATE by Papernot et al. \cite{DBLP:conf/iclr/PapernotAEGT17,DBLP:conf/iclr/PapernotSMRTE18} achieved DP, Cronos \cite{chang2019cronus} by Chang et al. protected ML from an MIA in a federated learning setting, and DMP \cite{shejwalkar2021membership} achieved a higher privacy-utility trade-off by removing public data with low entropy.

Currently, DMP is the best defense in the sense of the privacy-utility trade-off. However, it requires public data. Other known defenses, AdvReg and MemGuard, have an advantage in that they do not require public reference data.

\section{Preliminaries}
\subsection{Machine Learning}
\textit{An ML model} for a classification task is a function $F$ parameterized by internal \textit{model parameters}. It takes a $d$-dimensional real-valued vector $x\in\mathbb{R}^{d}$ as input and outputs a $c$-dimensional real-valued vector $\hat{y}=(\hat{y}_1,\ldots,\hat{y}_c)$. The output $\hat{y}$ has to satisfy $\hat{y}_i\in [0,1]$ and $\sum_i\hat{y}_i=1$. Each $\hat{y}_i$ is called a \textit{confidence score}. Its intuitive meaning is the likelihood of $x$ belonging to class $i$. The $\argmax_i\hat{y}_i$ is called a \textit{predicted label} (or \textit{predicted class}).

An ML model $F$ is trained using a \textit{training dataset} $D\subset\{(x,y) \mid x\in\mathbb{R}^{d},\ y\in\{0,1\}^{c}\}$, where $x$ is a data point, and $y$ is a one-hot vector reflecting the true class label of $x$. In the training procedure, the model parameters of $F$ are iteratively updated to reduce the predetermined \textit{loss} $\sum_{(x,y)\in D}L(F(x),y)$, which is the sum of errors between the prediction $F(x)$ and true label $y$. For inference, $F$ takes input $x$ and outputs $\hat{y}=F(x)$ as a \textit{prediction}.

The \textit{accuracy} of $F$ for dataset $D$ is the ratio between the number of elements $(x,y)\in D$ satisfying $\argmax_iF(x)_i=\argmax_iy_i$. Here, $F(x)_i$ and $y_i$ are the $i$-th component of $\hat{y}=F(x)$ and $y$, respectively. The \textit{training accuracy} and \textit{testing accuracy} of $F$ are for the training and testing datasets, respectively. Here, \textit{testing dataset} is a dataset that does not overlap with the training dataset. The \textit{generalization gap} of $F$ is the difference between training and testing accuracies. 

\subsection{Membership Inference Attack (MIA)}
\label{subsec:MIA-def}
MIA is an attack in which an attacker attempts to determine whether given data (called \textit{target data}) are used for training a given ML model (called a \textit{target model}). In the discussion of MIAs, the training data of the target model are called \textit{member data}, and non-training data are called \textit{non-member data}.

There are two types of MIAs, \textit{white-box} and \textit{black-box} \cite{shokri2017membership,DBLP:conf/sp/NasrSH19}. Attackers of the former can take as input the model structure and model parameters of the target model. Attackers of the latter do not take them as input but are allowed to make queries to the target model and obtain answers any number of times.
A black-box MIA can be divided into the two sub-types, \textit{MIA with confidence scores} and \textit{label-only MIA} \cite{choo2020label}. 
Attackers of the former can obtain confidence scores as answers from the target model but attackers of the latter can obtain only predicted labels as answers.

In all types of MIAs, the attackers can take the target data and \textit{prior knowledge} as inputs. Intuitively, the prior knowledge is what attackers know in advance. What type of prior knowledge an adversary can obtain depends on the assumed threat model. 
An example of prior knowledge is a dataset sampled from the same distribution as the training data of the target model, not overlapping with the training data. Another example is a portion of the training data. The prior knowledge we focused in this study is described in Section \ref{subsection:Setting-of-MIA}.
The \textit{attack accuracy} of an attacker for an MIA is the probability that they will succeed in inferring whether target data are member data. As in the all previous papers, the target data are taken from member data with a probability of $50\%$.

One of the main factor causing MIA risks is
\textit{overffiting} of an ML model on the training ($=$member) data.
The member data can be distinguished from non-member data \cite{yeom2018privacy,salem2019ml} depending on whether it is overfitted to the target model, e.g. by checking whether the highest confidence score of output of the target model is more than a given threshold.

\subsection{Distillation for Membership Privacy (DMP)}
\label{subsec:DMP}

\textit{DMP} \cite{shejwalkar2021membership} is a state-of-the-art defense method against MIAs that leverages knowledge distillation \cite{hinton2015distil}. Distillation was originally introduced as a model compression technique that transfers the knowledge of a large \textit{teacher model} to a small \textit{student model} by using the output of a teacher model obtained on unlabeled \textit{reference dataset}. DMP needs public reference dataset $R$ disjoint from the training dataset $D$ to train ML models with membership privacy.

The training algorithm of DMP is given in Algorithm \ref{alg:DMP}. Here, $L$ is the loss function. First, DMP trains a teacher model $F$ using a private training dataset $D$ (Step 1). $F$ is overfitted to $D$ and therefore vulnerable to MIA. Next, DMP computes the \textit{soft labels} $F(x)$ of each peice of data $x$ of public reference dataset $R$ and lets $\bar{R}$ be the set of $(x,F(x))$ (Step 2). Finally, to obtain a protected model, DMP trains a student model $H$ using the dataset $\bar{R}$ (Step 3). $H$ has MIA resistance because it is trained without direct access to the private $D$. Note that DMP uses $H$ with the same architecture as $F$.
\begin{algorithm}
\caption{Training algorithm of DMP}
\label{alg:DMP}
\begin{algorithmic}[1]
\REQUIRE  training dataset $D\subset\{(x,y)\mid x\in\mathbb{R}^{d}, y\in\{0,1\}^{c}\}$, reference dataset $R\subset\{x\mid x\in\mathbb{R}^{d}\}$, and initialized parameters of $F,H$. 
\ENSURE Distilled student model $H$.
\STATE Train $F$ by using $D$ as a training dataset until the training converges to minimize the loss $$\sum_{(x,y)\in D}L(F(x),y).$$
\STATE Let $\bar{R}$ be a dataset $R$ with soft labels, 
  $$\bar{R}=\{(x,F(x))\mid x\in R\}.$$
\STATE Train $H$ by using a dataset $\bar{R}$ until the training converges to minimize the loss 
$$\sum_{(x,y')\in\bar{R}} L(H(x),y').$$
\STATE Return $H$.
\end{algorithmic}
\end{algorithm}
The authors of DMP \cite{shejwalkar2021membership} proposed three different ways of achieving the desired privacy-utility tradeoffs:
\begin{itemize}
 \item increasing the temperature of the softmax layer of $F$,
 \item removing reference data with high entropy predictions from $F$,
 \item decreasing the size of the reference dataset.
\end{itemize}
All of the above changes reduce MIA risks but also the accuracy of $H$ and vice versa. 
When we use the second or third way to tune DMP, we select samples from the reference dataset and use them as $R$ in Step 2.

\section{Our Proposed Defense}

In this section, we propose a new defense that can protect ML from MIAs without using a reference dataset.

\subsection{Idea}
\label{subsection: Idea}
The starting point with our approach is DMP \cite{shejwalkar2021membership}.
That is, we train a \textit{teacher model} $F$ using a training dataset $D$, compute \textit{soft labels} $F(x)$ to $x$ 
of public \textit{reference dataset} $R$,  train a \textit{student model} $H$ using $(x,F(x))$, and, finally, use $H$ for inference. DMP can mitigate MIAs as described in Section \ref{subsec:DMP}.

The problem with DMP is that it requires a public reference dataset, which may be difficult to collect in privacy-sensitive domains \cite{shejwalkar2021membership}. 
A na\"{i}ve idea to solve this problem is to use the original $D$ as a reference dataset. However, our experiment shows that this approach does not sufficiently mitigate the MIA risk (see Section \ref{subsec:Discussions}).
The main problem of the na\"{i}ve idea is that data $x$ of the reference dataset $R=D$ is \textit{member data} of $F$. Therefore, $F$ results in overfitting on $x$ and the confidence score $\hat{y}=F(x)$ is close to the one-hot vector $y$ of the true label. Hence, $H$ trained on $(x,\hat{y})$ results again in overfitting on $x$, which can be exploited by an MIA.

Our proposed defense, denoted by \textit{knowledge cross-distillation} (KCD) is designed to overcome the above problem. We divide the training dataset into $n$ parts, leave one part as a reference dataset, and train a teacher $F_1$ using the remaining parts. To increase the accuracy of KCD, we prepare teachers $F_2,\ldots,F_n$ as well and repeat the above procedure for each teacher by changing the reference part. Finally, we use each reference part to distill the knowledge of  each corresponding teacher into a single $H$.
Our defense solves the problem of the na\"{i}ve idea because none of the remaining parts of the training dataset are used to train the teacher model.

\subsection{Description}
\label{subsection:Description}

\begin{figure}
	\centering
	\includegraphics[width=7.5 cm]{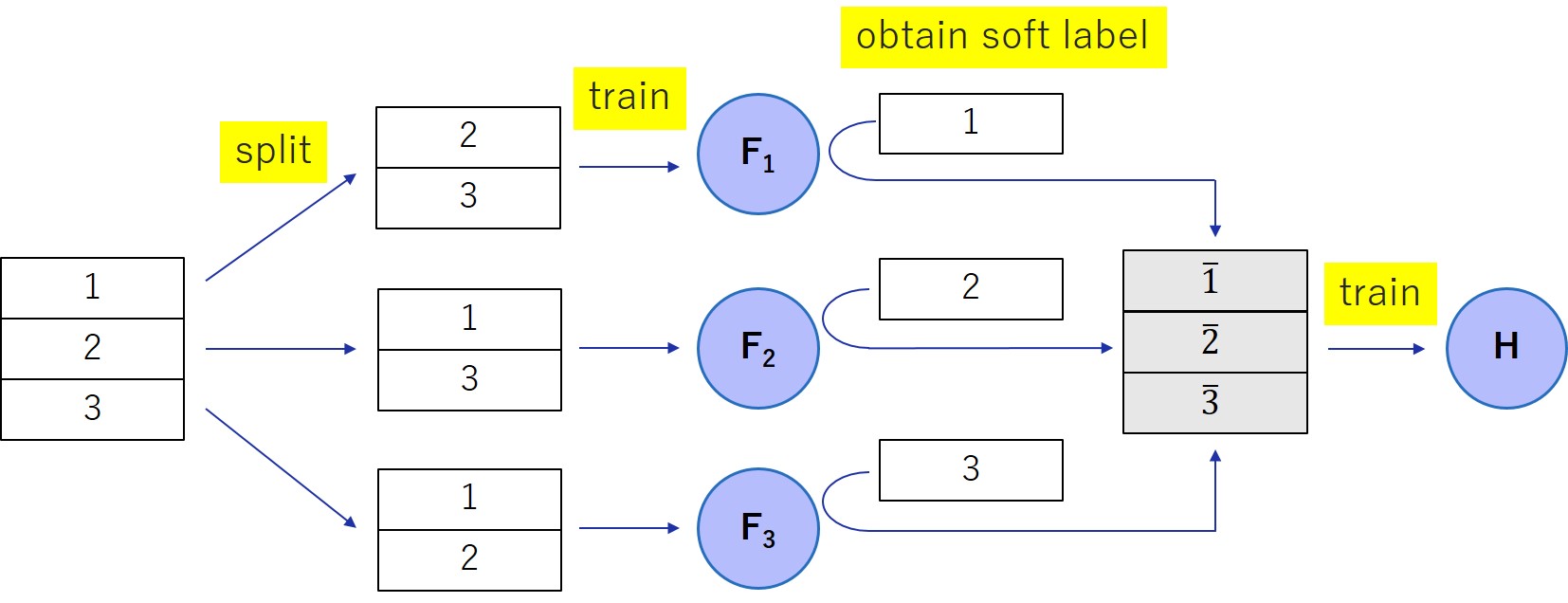}
	\caption{\rm Outline of KCD when dividing training dataset into three subsets. $F_{1}$-$F_{3}$: teacher models, H: student model.}\label{fig:outline}
\end{figure}

The training algorithm of the our proposed defense, KCD,  is given in Algorithm \ref{alg:MDMP} and is overviewed  in Figure \ref{fig:outline}. Here, $F_1,\ldots,F_n$, and $H$ are models with the same structure as that of the model $F$ that we want to protect\footnote{Although we use the term ``distillation,'' we use teacher and student models with the same structure as in DMP \cite{shejwalkar2021membership}. This is because we are not concerned about the size of the resulting model.}. $L$ is the loss function.

In Algorithm \ref{alg:MDMP}, we divide training dataset $D$ into $n$ disjoint 
subsets $D_1,\ldots,D_{n}$ with almost the same size, such that $D=\bigsqcup^{n}_{i=1}D_i$ holds\footnote{$\bigsqcup^{n}_{i=1}D_i$ denotes a disjoint union of sets} (Step 1).
Then, for $i=1,\ldots,n$, we train the teacher model $F_i$ using the dataset $D$ but exclude $D_i$ (Step 2-4). Let $\bar{D}_i$ be the dataset that is obtained by adding soft labels $F_i(x)$ to $(x,y)\in D_i$ (Step 5). Finally, we train a student model $H$ using the dataset $\bigcup_i\bar{D}_i$ to minimize the combined loss function with hyperparameter $\alpha$ (Step 6). 

Our loss function comprises two terms; the first term is the loss for soft labels $y'$, and the second is the loss for the true label $y$. The hyperparameter $\alpha$ can tune the privacy-utility trade-off of KCD. In fact, if $\alpha=1$, our defense protects the privacy of the training data due to the reason mentioned in Section \ref{subsection: Idea}. If $\alpha=0$, KCD  becomes the same as the unprotected ML.

Note that our privacy-utility trade-off based on $\alpha$ cannot be directly applied to the known knowledge distillation-based defenses, DMP \cite{shejwalkar2021membership} and Cronus \cite{chang2019cronus}, because the public reference datasets for these defenses do not have the true labels and loss for the predicted scores and true labels cannot be computed.

\begin{algorithm}
\caption{Training algorithm of KCD}
\label{alg:MDMP}
\begin{algorithmic}[1]
\REQUIRE  training dataset $D\subset\{(x,y)\mid x\in\mathbb{R}^{d}, y\in\{0,1\}^{c}\}$, hyperparameter $\alpha\in[0,1]$, and initialized parameters of $F_1,\ldots,F_{n},H$. 
\ENSURE Distilled student model $H$.
\STATE Divide $D$ into $n$ randomly selected disjoint subsets $\{D_i\}^{n}_{i=1}$ with almost the same size, such that\footnotemark[3] $$D=\bigsqcup^{n}_{i=1}D_i, $$ 
\FOR{$i=1,\ldots,n$}
\STATE Train $F_{i}$ by using $D\setminus D_i$ as a training dataset until the training converges to minimize the loss $$\sum_{(x,y)\in D\setminus D_i}L(F_i(x),y).$$
\ENDFOR
\STATE Let $\bar{D}_i$ be a dataset $D_i$ with soft label 
  $$\bar{D}_i=\{(x,F_i(x))\mid \exists y~:~(x,y)\in D_i\},$$
  and let $\bar{D}=\cup_i\bar{D}_i$.
\STATE Train $H$ by using a dataset $\bar{D}$ until the training converges to minimize the loss 
    \begin{equation}\label{mergedLoss}
        \alpha\sum_{(x,y')\in\bar{D}} L(H(x),y')+(1-\alpha) \sum_{(x,y)\in D}L(H(x),y)
    \end{equation}
\STATE Return $H$.
\end{algorithmic}
\end{algorithm}

\section{Experimental Setup}

We conducted our experiments using the following datasets and model architectures as in the previous studies \cite{shokri2017membership, nasr2018machine, jia2019memguard, choo2020label, DBLP:conf/sp/NasrSH19, shejwalkar2021membership, song2020systematic}.

\subsection{Datasets}
\textbf{CIFAR 10:} This is a typical benchmark dataset used for evaluating the performance of image-classification algorithms \cite{Krizhevsky09}. It contains $60,000$ RGB images. Each image is composed of $32 \times 32$ pixels and labeled in one of $10$ classes.

\vspace{1em}\noindent\textbf{Purchase100:}
This is a benchmark dataset used for MIAs.
It is based on a dataset provided by Kaggle's Acquire Valued Shoppers Challenge \cite{Purchase100}.
We used a processed and simplified one by Shokri et al. \cite{shokri2017membership}.
The dataset has $197,324$ records with $600$ binary features, each of which represents whether the corresponding customer purchased an item. The data are clustered into $100$ classes representing different purchase styles, and the classification task is to predict which one of the $100$ classes an input is in. 

\vspace{1em}\noindent\textbf{Texas100:} This is also a benchmark dataset used for MIAs. It is based on the hospital discharge data \cite{Texas100} from several health facilities published by the Texas Department of State Health Services and was processed and simplified by Shokri et al. \cite{shokri2017membership}. It contains the 100 most frequent procedures that patients underwent. The dataset has $67,330$ records with $6,170$ binary features of patients, such as the corresponding patient's symptoms and genetic information. The classification task is to predict which one of the $100$ procedures a patient for a piece of an input data underwent.

\subsection{Model Architectures}

\noindent\textbf{Wide ResNet-28:} For CIFAR 10, we used the same model architecture as in a previous study \cite{choo2020label}, i.e., Wide ResNet-28. 

\vspace{1em}\noindent\textbf{Purchase and Texas classifiers:} For Purchase 100 and Texas 100, we used fully connected NNs with Tanh activation functions. We used the same layer sizes as in a previous study \cite{nasr2018machine}, i.e., layer sizes $(1024, 512, 256, 128)$.

\subsection{Setting of MIA}
\label{subsection:Setting-of-MIA}
As in the previous studies of MIAs \cite{nasr2018machine,DBLP:conf/sp/NasrSH19}, we consider a strong setting where the attackers know the non-member dataset and a subset of the member dataset of the target model as prior knowledge. 
(This subset of the member dataset does not contain the target data, of course).
This setting is called \textit{supervised inference} \cite{DBLP:conf/sp/NasrSH19}.

One may think that the supervised inference setting seems too strong as a real setting. However, the \textit{shadow model} technique \cite{shokri2017membership} allows an attacker to achieve supervised inference virtually \cite{DBLP:conf/sp/NasrSH19}. A shadow model is an ML model that is trained by an attacker to mimic a target ML model. The attacker then knows the training data of the shadow model as in the supervised inference setting since the attacker trains it.

\subsection{MIAs for Evaluations}
We conducted comprehensive experiments for three types of MIA: black-box MIA with confidence score, black-box MIA with only labels, and white-box MIA.

\subsubsection{Black-box MIA with confidence score (BB w/score)}

These are attacks such that the attackers know the confidences scores as outputs of the target model. There are two sub-types of these attacks.

\vspace{1em}\noindent\textbf{NN-based attack:} This is a type of black-box MIA using an NN, called \textit{attack classifier} $A$. Specifically, the attacker knows a set of non-member data and a subset of member data as their prior knowledge, as mentioned in Section \ref{subsection:Setting-of-MIA}. They send these data to the target model and obtain their confidence scores as answers. Using these data, the answers, and the knowledge of whether these data are members, they train $A$. Finally, they infer the membership status of the target data by taking their label and confidence score as input to $A$.
There are two known NN-based attacks \cite{shokri2017membership,salem2019ml}. The difference between them is whether the attacker trains an attack classifier for each label class; the original attack by Shokri et al. \cite{shokri2017membership} uses one classifier per each class and a simplified attack by Salem et al. \cite{salem2019ml}, called \textit{ML Leaks Adversary 1}, uses only one common attack classifier for all classes.

{In our experiments,} we executed the attack \textit{ML Leaks Adversary 1} \cite{salem2019ml} since it is simpler and ``has very similar membership inference'' \cite{salem2019ml} to that of Shokri et al. \cite{shokri2017membership}. 

\vspace{1em}\noindent\textbf{Metric-based attack:} This is a type of black-box MIA that directly uses the fact that the confidence score $F(x)$ of the target data $(x,y)$ differs depending on whether $(x,y)$ is a member. Specifically, an attacker computes a value $m=M(F(x),y)$, called a \textit{metric}, and infers $(x,y)$ as a member if $m$ satisfies a given condition (e.g., greater than a given threshold).
{There are five known attacks of this type: \textit{Top 1}, \textit{correctness}, \textit{confidence}, \textit{entropy}, and \textit{m-entropy attacks} (Table \ref{tab:metric-based}),
where Top 1 was proposed in \cite{salem2019ml}, and the other four were proposed in \cite{song2020systematic} by generalizing or improving known metric-based attacks \cite{yeom2018privacy,DBLP:conf/ccs/SongSM19,salem2019ml,DBLP:conf/uss/LeinoF20,salem2019ml,song2020systematic}.} 

{In our experiments,} we executed all five metric-based attacks \cite{salem2019ml,song2020systematic} mentioned above.

\begin{table}[t]
 \centering
 \begin{tabular}{l||l}
 \hline
 \rm Name & \rm Condition \\
 \hline\hline
 \rm Top 1 &   $\argmax_iF(x)_i\overset{?}{\ge} \tau$\\
  \hline 
 \rm Correctness & $\argmax_iF(x)_i\overset{?}{=}\argmax_iy_i$\\
 \hline 
 \rm Confidence & $F(x)_{\ell[y]} \overset{?}{\ge} \tau_{\ell[y]}$\\
 \hline
 \rm Entropy & $-\sum_i F(x)_i\log F(x)_i \overset{?}{\le} \tau_{\ell[y]}$\\
 \hline
 \rm Modified Entropy & $\begin{array}{l}
       -(1-F(x)_{\ell[y]})\log F(x)_{\ell[y]}\\
       ~~~-\sum_{i\neq \ell[y]} F(x)_i\log F(x)_i
       \overset{?}{\le} \tau_{\ell[y]}\\
 \end{array}$\\
 \hline
 \end{tabular}
   \caption{\rm {Known metric-based attacks. Here, $F(x)_i$ and $y_i$ mean $i$-th component of $F(x)$ and $y$ respectively and $\ell[y]$ is the label corresponding to one-hot vector $y$, that is, $\operatorname{argmax}_i y_i$. $\tau$ and $\tau_{\ell[y]}$ are thresholds determined by attackers.}}
   \label{tab:metric-based}
 \end{table}

\subsubsection{Black-box MIA only with labels (BB label only)}

{These are attacks such that} attackers know only the predicted labels as outputs of the target model without knowing the confidence scores. We call such an MIA a \textit{label-only MIA}.
There are two known label-only attacks, \textit{boundary distance (BD)} and \textit{data augmentation} \cite{choo2020label}. {We introduce only the former one} because it is stronger than the {latter one} \cite{choo2020label}.

A BD attack is an attack that {computes the smallest} adversarial perturbation {$\Delta x$} satisfying $\argmax_iF(x+\Delta x)_i\neq \argmax_iy_i$ for the target data $(x,y)$. Here, $F(x+\Delta x)_i$ and $y_i$ are the $i$-th components of $F(x+\Delta x)$ and $y$, respectively. The attacker then infers $x$ is a member if the $L_2$ norm of $\Delta x$ is larger than a predetermined threshold. 
A BD attack {is} a black-box {MIA} if adversarial perturbation is crafted by HopSkipJump \cite{chen2020hopskipjumpattack}. However, the attack becomes a white-box MIA if we use the Carlini-Wagner {method for} adversarial perturbation \cite{DBLP:conf/sp/Carlini017}.

In our experiments, we executed the BD attack with HopSkipJump. This is because the attack accuracy of the BD attack based on HopSkipJump is asymptotically equal to that of the BD attack with Carlini-Wagner when the number of queries increases \cite{choo2020label}.

\subsubsection{White-box MIA (WB)}
{These are attacks such that attackers can take the confidence score of target data besides the model structure and model parameters of the target model as input.}
Two white-box attacks have been proposed, the \textit{Nasr-Shokri-Houmansadr (NSH) attack} \cite{DBLP:conf/sp/NasrSH19} and \textit{Hui's attack} \cite{DBLP:conf/ndss/HuiYYBGC21}.

The NSH attack exploits the fact that the gradient for the model parameter of the target model $F$ on $(x,y)$ becomes smaller if $(x,y)$ is a member of $F$. Specifically, an attacker computes the gradient of $F$ on the target data and infers the membership of $(x,y)$ by inputting the gradient {as well as the confidence score and the class label} into an NN trained by the attacker. 
{Hui's attack} focuses mainly on reducing the assumption behind the NSH attack. That is, it can be executed without assuming that an attacker has member data as prior knowledge \cite{shokri2017membership}.

{In our experiments,} we executed only the NSH attack, {since} our assumption was stronger than Hui's; an attacker has member data as prior knowledge (as mentioned in Section \ref{subsection:Setting-of-MIA}).

\subsection{Known Defenses}
\label{subsection:Known_Defenses}

Known defenses can be categorized into the following three types.
We chose the best defense from all three types for comparison with our method.

\vspace{1em}\noindent\textbf{Regularization-based methods:}
These methods use the fact that the regularization techniques of ML models mitigate overfitting, one of the main reasons behind the MIA risk \cite{yeom2018privacy}. Regularization techniques, such as $L_2$-regularization, dropout \cite{DBLP:journals/jmlr/SrivastavaHKSS14}, and early-stopping, also mitigate the MIA risk, as pointed out by Nasr et al. \cite{nasr2018machine}, Shokri et al. \cite{shokri2017membership}, and Song et al. \cite{song2020systematic}, respectively.
{Meanwhile,} \textit{Adversarial Regularization} (AdvReg) \cite{nasr2018machine} is a regularization that is {focused on} mitigating MIAs.
To conduct our experiment, we chose AdvReg from this type of attack since it mitigates the MIA risk {the best}. 

AdvReg is based on a game theoretic framework similar to GANs \cite{HayesMDC19LOGAN}. Specifically, we train a {model} $F$ we want to protect and a {pseudo attacker} $A$ alternatively. The aim of the $A$ is to distinguish member data from non-member data. It corresponds to a discriminator of a GAN, and the gain of the $A$ is added to the loss of $F$ as a regularization term.

\vspace{1em}\noindent\textbf{AX (Adversarial eXample)-based method:}
This method exploits an AX \cite{DBLP:journals/corr/SzegedyZSBEGF13} to mitigate the MIA risk, {where AX is a technique for deceiving ML by adding small noise to the input of the ML.}
We used \textit{MemGuard} \cite{jia2019memguard} in our experiments.
MemGuard adds AX noise to the output of $F$, which we want to protect. Then, an attacker who uses an NN to attack $F$ is deceived by the noise and cannot accurately determine the membership of the target data. 

\vspace{1em}\noindent\textbf{KT (Knowledge Transfer)-based methods:}
These methods exploit KT to mitigate the MIA risk. Here, KT means knowledge distillation (explained in Section \ref{subsection: Idea}) or its variants. There are three known KT-based methods: {\textit{DMP} \cite{shejwalkar2021membership}, \textit{PATE} \cite{DBLP:conf/iclr/PapernotAEGT17}, and an improved variant of PATE, \textit{PATE with confident-GNMax} \cite{DBLP:conf/iclr/PapernotSMRTE18}.} We used DMP and PATE with confident-GNMax {in our experiments} for image data. However, we used only DMP for tabular data because PATE with confident-GNMax requires GANs.

Details on DMP have already been given in Section \ref{subsec:DMP}. Meanwhile, PATE trains multiple teacher models with \textit{disjoint} subsets of private training data, gives public data hard labels chosen by noisy voting among the {teachers}, and finally trains a student model using labeled public data. {A noisy voting mechanism provides differential privacy guarantees with respect to the training data.} Confident-GNMax is a new noisy aggregation method for improving the original PATE. 
{To achieve a smaller privacy budget $\varepsilon$, instead of labeling all public data, it} selects the samples among public data to be labeled by checking if the result of a {noisy} plurality vote crosses a threshold. {Once the threshold and noise parameters are determined, $\varepsilon$ can be computed. We train a student model using semi-supervised learning with GANs \cite{NIPS2016_8a3363ab}.}

\subsection{ML Setups and Hyperparameter Choosing}

\subsubsection{ML setups}
\label{subsection:ML Setups}
\label{app:datasplit}

In all experiments, we used a batch size of 64, the SGD optimizer with a momentum of $0.9$ and weight decay of $10^{-5}$, and the ReduceLROnPlateau scheduler with default hyperparameters.
The model that recorded the best validation accuracy in five trials was evaluated to test the accuracy and risks against the four types of MIAs.
We conducted all experiments using the PyTorch 1.7 framework on a Tesla V100 GPU with 32-GB memory.

\begin{table*}[t]
  \centering
  \begin{tabular}{|l||r|r|r|r|r|r|r|r}
    \hline
     \multicolumn{1}{|c||}{ } & \multicolumn{3}{c|}{\rm Train.}  & \multicolumn{1}{c|}{\rm Ref.} & \multicolumn{1}{c|}{\rm Val.} & \multicolumn{3}{c|}{\rm Test.} \\ \cline{1-4}\cline{7-9}
    \multicolumn{1}{|c||}{\rm Dataset} & \rm All & \rm Known & \rm Target &   &   & \rm  All & \rm  Known &   \multicolumn{1}{c|}{\rm Target} \\ \hline
    \multicolumn{1}{|c||}{\rm Purchase} & \rm  10000 & \rm  5000 & \rm  2500 & \rm  10000 & \rm  5000 & \rm  5000 & \rm  2500 &   \multicolumn{1}{c|}{\rm 2500} \\ \hline
    \multicolumn{1}{|c||}{\rm Texas} & \rm  10000 & \rm  5000 & \rm  2500 & \rm  10000 & \rm  5000 & \rm  5000 & \rm  2500 & \multicolumn{1}{c|}{\rm 2500} \\ \hline
    \multicolumn{1}{|c||}{\rm CIFAR10} & \rm  25000 & \rm  12500 & \rm  2500 & \rm  25000 & \rm  5000 & \rm  5000 & \rm  2500 & \multicolumn{1}{c|}{\rm 2500} \\ \hline
  \end{tabular}
  \caption{Dataset splits. \rm ``All'': All data used to train or test, ``Known'': Known data that attacker can exploit to execute MIA, ``Target'': Target data for which attacker attempts to infer membership.}\label{table:dataset_splits}
\end{table*}

Table \ref{table:dataset_splits} shows how we split the above datasets in our experiments.
{Here, \textit{validation dataset} is a dataset used to select the best model parameters that does not overlap with the training dataset in our experiment.}
Following {the} previous studies \cite{nasr2018machine,DBLP:conf/sp/NasrSH19}, we considered strong attackers who know the non-member dataset and a subset of the member dataset of the target model as their prior knowledge (see Section \ref{subsection:Setting-of-MIA}). We used the rest of the training/testing data as the target data to execute an MIA. The amounts of known data and target data are also depicted in Table~\ref{table:dataset_splits}.

\subsubsection{Hyperparameter tuning}\label{app:tuning}

\noindent\textbf{Unprotected, AdvReg, MemGuard, DMP, and KCD:}
Using Optuna \cite{Optuna}, 
we optimized hyperparameters for each scheme.
\begin{itemize}
 \item For unprotected models, we chose learning rates that maximize validation accuracies.
 \item For AdvReg, MemGuard, DMP, and KCD, we tuned the learning rates and their specific parameters, i.e., the penalty parameter $\lambda$ (AdvReg), learning rate $\beta$ of a pseudo attacker, the weights $c_2$, $c_3$ of the loss function (MemGuard), the size of the public reference data\footnote{There are three privacy-utility trade-off hyperparameters depicted in the DMP paper \cite{shejwalkar2021membership}, temperature, entropy criterion, and the number of reference data {as explained in Section \ref{subsec:DMP}}. We chose the number of pieces of reference data from them for our experiments since this number shows the best trade-off in our environment.} (DMP), and the intensity $\alpha$ of the distillation in Algorithm \ref{alg:MDMP} (KCD), respectively.
 We optimized their hyperparameters toward a high validation accuracy and low MIA risk.
\end{itemize}

The hyperparameters of the defenses were basically chosen to have almost the same accuracy as the unprotected model and a considerably low MIA risk, except for some defenses whose accuracy inevitably drops no matter which hyperparameters we chose for them with low MIA risk.
\begin{itemize}
 \item In Tables \ref{table:comparison-purchase}, \ref{table:comparison-texas}, and \ref{table:comparison-CIFAR10},
 we chose hyperparameters for AdvReg, that enable a better privacy-utility tradeoff (i.e., relatively small validation accuracy drop and mid-level MIA resistance)
 since almost the same validation accuracy as the unprotected model (making the MIA risk similar to a random guess, resp.) results in an MIA risk that is the same as that of an unprotected model (deterioration of validation accuracy, resp.).
 \item For MemGuard, we fixed $\varepsilon=1.0$ and tuned the other parameters toward a low MIA risk.
 \item The hyperparameters of DMP were chosen to replicate the performance of the original paper \cite{shejwalkar2021membership}.
 \item For KCD, we chose a model whose validation accuracy is close to that of DMP.
\end{itemize}

\noindent\textbf{PATE:}
For PATE, we trained four ensembles of teachers, i.e., 3, 5, 10 and 25, and selected five different privacy levels $(\varepsilon,\delta)$ for each ensemble (where $\delta$ is fixed to $10^{-4}$ as the order of the size of the public reference dataset is $10^4$ \cite{DBLP:conf/iclr/PapernotSMRTE18}). Since our interest is empirical MIA resistance, not DP guarantees, we chose various values for $\varepsilon$. For example, for three teachers, we chose $\varepsilon=229, 1473, 6291, 36849, 83535,141923$ (These cannot be round numbers because these values are automatically computed after we choose the thresholds and noise parameters). Epsilon $141923$ was the minimum value that maximized the validation accuracy (i.e., corresponding to the non-private case), and epsilon $229$ was the minimum value that provided enough labeled public data to train a student. Using Optuna, we optimized the learning rates toward a high validation accuracy for each $\varepsilon$.

In Table 5, we chose the hyperparameters ``$3$ teachers, $\varepsilon=141923$, $\delta=10^{-4}$,'' which maximized the validation accuracy because all the trained  models had almost the same MIA risks.

\subsubsection{Choice of loss function}\label{app:loss}

The loss functions for most of the defenses were chosen from the original studies.
The exception is DMP with synthetic reference data \cite{shejwalkar2021membership}; we chose the mean squared error (MSE) as a loss function since ``synthetic'' DMP with this loss function performed better than the original loss function, KL divergence, in our experiments.
We chose a suitable loss function on the basis of known facts about distillation: the KL loss at a high temperature $T$ asymptotically approaches the MSE, and which of these performs well is an empirical question \cite{hinton2015distil} (the loss with $T=1$ is KL loss).
Therefore, we examined the KL divergence loss at various $T$ for ``synthetic'' DMP and found that a higher $T$ leads to better performance and that MSE loss is the best.
By doing similar experiments for our defense, KCD, we determined the suitable loss to be the MSE for the Purchase100 and CIFAR10 datasets and KL divergence with $T=1$ for the Texas100 dataset.

\subsubsection{Notes on implementation of DMP}

The published code\footnote{https://github.com/vrt1shjwlkr/AAAI21-MIA-Defense} does not include reference data selection but nonetheless achieved good results.
Therefore, we did not implement entropy-based criteria for DMP in our experiments.

For DMP with synthetic reference data, we trained (unconditional) DCGAN as in the original study \cite{shejwalkar2021membership}.
We trained them to obtain generated images in accordance with the implementation of PyTorch  examples\footnote{\url{https://github.com/pytorch/examples/blob/master/dcgan/main.py}}.
Since the resulting images (Figure~\ref{fig:generated_cifar10}) were natural and showed large diversity, we considered them to be sufficient for the reference dataset of DMP.

\begin{table*}
\begin{center}
\begin{tabular}{|l|l|l||r|r|r|r|r|}
    \hline
         \multicolumn{1}{|c|}{ }&\multicolumn{2}{c||}{\multirow{2}{*}{\rm Defense}}& \multicolumn{5}{c|}{\rm Purchase100} \\ 
         \cline{4-8}
         \multicolumn{1}{|c|}{ }&\multicolumn{2}{c||}{ }
         &
         \multicolumn{1}{c|}{\multirow{2}{*}{\rm Train}}&
         \multicolumn{1}{c|}{\multirow{2}{*}{\rm Test}} & \multicolumn{2}{c|}{\rm BB} & \multicolumn{1}{c|}{\multirow{2}{*}{\rm WB}}\\ 
         \cline{2-3}\cline{6-7}
         \multicolumn{1}{|c|}{ }& \rm Category & \rm Name &&
          &  \rm w/score & \rm label only &\multicolumn{1}{c|}{~} \\ 
        \hline\hline
        \multicolumn{1}{|c|}{ }&\rm Reg-based&\rm AdvReg \cite{nasr2018machine} &\rm 82.3\%& 
        \rm 64.2\%  &\rm 59.9\% &\rm 58.9\%&\multicolumn{1}{c|}{\rm 60.2\%}\\ 
        \cline{2-8}
        \multicolumn{1}{|c|}{\rm ${}^\nexists$Public Ref.}
        &\rm AX-based&\rm MemGuard \cite{jia2019memguard} & \rm \bf 100.0\%&
        \bf 77.0\% &\rm 72.1\% &\rm (68.6\%)&\multicolumn{1}{c|}{\rm (74.3\%)} \\ 
        \cline{2-8}
        \multicolumn{1}{|c|}{ }&\rm KT-based& \rm \bf KCD & \rm 93.8\%&
        \rm 75.7\%  & \bf 58.8\% &\bf 58.7\%& \multicolumn{1}{c|}{\bf 59.5\%}  \\
        \hline\hline
        \multicolumn{1}{|c|}{\rm ${}^\exists$Public Ref.}&
        \rm KT-based&\rm DMP \cite{shejwalkar2021membership}\hspace{6em}  &
        \rm 89.3\%&
        \rm 75.4\% &\rm 57.1\% &\rm 57.5\%&\multicolumn{1}{c|}{\rm 57.3\%} \\
        \hline\hline
        \multicolumn{3}{|c||}{\rm Unprotected} &
        \rm 100.0\%& 
        \rm 77.0\%  &\rm 73.7\%  &\rm 68.6\%&\multicolumn{1}{c|}{\rm 74.3\%} \\ 
        \hline
    \end{tabular}
    \caption{\centering Comparisons on Purchase100.}\label{table:comparison-purchase}
    \begin{tabular}{|l|l|l||r|r|r|r|r|}
    \hline
         \multicolumn{1}{|c|}{ }&\multicolumn{2}{c||}{\multirow{2}{*}{\rm Defense}} &\multicolumn{5}{c|}{\rm Texas100}  \\ 
         \cline{4-8}
         \multicolumn{1}{|c|}{ }&\multicolumn{2}{c||}{ }
         &\multicolumn{1}{c|}{\multirow{2}{*}{\rm Train}}&
         \multicolumn{1}{c|}{\multirow{2}{*}{\rm Test}} & \multicolumn{2}{c|}{\rm BB} & \multicolumn{1}{c|}{\multirow{2}{*}{\rm WB}}  \\ 
         \cline{2-3}\cline{6-7}
         \multicolumn{1}{|c|}{ }& \rm Category & \rm Name &
          & 
          &  \rm w/score &\rm label only & \multicolumn{1}{c|}{ }\\ 
        \hline\hline
        \multicolumn{1}{|c|}{ }&\rm Reg-based&\rm AdvReg \cite{nasr2018machine} & \rm 60.5\%&
        \rm 45.5\%  &\rm 59.5\% &\rm 56.7\%& \multicolumn{1}{c|}{\rm 58.0\%} \\ 
        \cline{2-8}
        \multicolumn{1}{|c|}{\rm ${}^\nexists$Public Ref.}
        &\rm AX-based&\rm MemGuard \cite{jia2019memguard} &\bf 90.7\%&
        \bf 52.5\% &\rm 68.6\% &\rm (69.7\%) &\multicolumn{1}{c|}{\rm (70.0\%)}\\ 
        \cline{2-8}
        \multicolumn{1}{|c|}{ }&\rm KT-based& \rm \bf KCD & 
        \rm 59.2\% &\rm 52.0\%  &\bf 56.2\%&\bf 53.6\%& \multicolumn{1}{r|}{\bf 55.8\%}\\
        \hline\hline
        \multicolumn{1}{|c|}{\rm ${}^\exists$Public Ref.}&
        \rm KT-based&\rm DMP \cite{shejwalkar2021membership}\hspace{6em} &
        \rm 65.1\%&
        \rm 51.9\% &\rm 56.3\% &\rm 56.1\%& \multicolumn{1}{c|}{\rm 56.5\%}\\
        \hline\hline
        \multicolumn{3}{|c||}{\rm Unprotected} &
        \rm 90.7\%&
        \rm 52.5\% &\rm 69.9\%  &\rm 69.7\%& \multicolumn{1}{c|}{\rm 70.0\%}\\ 
        \hline
    \end{tabular}
    \caption{\centering Comparisons on Texas100.}\label{table:comparison-texas}
    \begin{tabular}{|l|l|l||r|r|r|r|r|}
    \hline
     \multicolumn{1}{|c|}{ }&\multicolumn{2}{c||}{\multirow{2}{*}{\rm Defense}}&\multicolumn{5}{c|}{ \rm CIFAR10}\\
    \cline{4-8}
        \multicolumn{1}{|c|}{ }&\multicolumn{2}{c||}{ }& \multicolumn{1}{c|}{\multirow{2}{*}{\rm Train}} & \multicolumn{1}{c|}{\multirow{2}{*}{\rm Test}}&\multicolumn{2}{c|}{\rm BB}&\multicolumn{1}{c|}{\multirow{2}{*}{\rm WB}} \\
        \cline{2-3}\cline{6-7}
        \multicolumn{1}{|c|}{ } &\multicolumn{1}{c|}{\rm Category}&\multicolumn{1}{l||}{\rm Name} &&
          &\rm w/score &\rm label only &\multicolumn{1}{c|}{ }\\ 
        \hline\hline
        \multicolumn{1}{|c|}{\multirow{4}{*}{${}^\nexists$\rm Public Ref.}}&\rm Reg-based &\rm AdvReg \cite{nasr2018machine} &\rm 84.9\%&\rm 
        76.3\%&\rm 54.6\% &\rm 
        54.7\%& 
        \multicolumn{1}{|c|}{\rm 55.2\%}\\ 
        \cline{2-8}
        \multicolumn{1}{|c|}{}
        &\rm AX-based &\rm MemGuard \cite{jia2019memguard} &\bf 100.0\% &
        \rm 82.1\%&\rm 64.3\% &
        \rm (55.6\%)&
        \multicolumn{1}{|c|}{\rm (66.0\%)}\\ 
        \cline{2-8}
        \multicolumn{1}{|c|}{ }&\multirow{2}{*}{\rm KT-based} & \rm DMP \cite{shejwalkar2021membership} (synth. ref.) &\rm 81.1\%&
        \rm 75.5\%&\bf 52.5\% &
        \bf 52.5\%&\multicolumn{1}{|c|}{\bf 52.6\%}\\ 
        \cline{3-8}
        \multicolumn{1}{|c|}{ }&& \bf KCD &\rm 94.0\% &
        \bf 82.2\%  &\rm 55.8\% &
        \rm 55.6\%&
        \multicolumn{1}{|c|}{\rm 56.2\%}\\
        \hline\hline
        \multicolumn{1}{|c|}{\multirow{2}{*}{\rm ${}^\exists$Public Ref.}}&\multirow{2}{*}{\rm KT-based} & \rm DMP \cite{shejwalkar2021membership} &\rm 84.2\% &
        \rm 82.2\% &\rm 51.1\% &
        \rm 50.9\% &
        \multicolumn{1}{|c|}{\rm 51.4\%} \\
        \cline{3-8}
        \multicolumn{1}{|c|}{ }&& \rm PATE  \cite{DBLP:conf/iclr/PapernotSMRTE18}&\rm  74.2\%&
        \rm 72.8\%  & \rm 51.2\% &
        \rm 50.2\% & 
        \multicolumn{1}{|c|}{\rm 51.4\%}\\
        \hline\hline
        \multicolumn{3}{|c||}{\rm Unprotected} &\rm 100.0\%& 
        \rm 82.1\%&\rm 65.9\% &
        \rm 65.4\%&\multicolumn{1}{|c|}{\rm 66.0\%}\\ 
        \hline
    \end{tabular}
    \caption{\centering Comparisons on CIFAR10}\label{table:comparison-CIFAR10}
\end{center}
{\sf Explanatory notes on above three tables:}
\small
\begin{itemize}
    \item Rows
    \begin{itemize}
    \item ``${}^\nexists$Public Ref.'' (resp. ``${}^\exists$Public Ref.'') means defense methods for ML models not using (resp. using) public reference data.  
    \item The bold means the best value in each column among the defenses of 
    ``${}^\nexists$Public Ref.''
    \item In Table \ref{table:comparison-CIFAR10}, ``DMP \cite{shejwalkar2021membership} (synth. ref.)'' is DMP with public reference data generated using DCGAN \cite{dcgan}.
    \item In Table \ref{table:comparison-CIFAR10}, ``PATE'' is PATE with confident-GNMax \cite{DBLP:conf/iclr/PapernotSMRTE18}.
    \end{itemize}
    \item Columns
    \begin{itemize}
        \item ``Train'' and ``Test'' are the training and testing accuracies.
        \item ``BB w/score'' is the maximum attack accuracies of the following black-box MIAs using confidence scores, ML Leaks Adversary 1 \cite{salem2019ml} and five metric-based attacks \cite{salem2019ml,song2020systematic}. 
        See Appendix \ref{section:Missing_Details_of_Experimental_Results} for the attack accuracy of each attack. 
        \item ``BB label only'' and ``WB'' are the attack accuracies of the BD attack \cite{choo2020label} and 
        NSH attack \cite{DBLP:conf/sp/NasrSH19}, respectively. 
    \end{itemize}
    \item Others
    \begin{itemize}
        \item The values of MemGuard in ``BB label only'' were not obtained in the experiments. We included the same values as the unprotected models because, as explained in \cite{choo2020label}, MemGuard works in such a manner that a model's predicted labels are not changed using the defense; therefore, the attack accuracies of label-only attacks for MemGuard become the same as those of unprotected models.
        \item Similarly, we included the same values as the unprotected models for MemGuard in ``WB'' because MemGuard is designed for black-box MIAs, and attackers using white-box MIAs can easily recover an unprotected model from MemGuard.
    \end{itemize}
\end{itemize}
\end{table*}

\begin{figure*}[htbp]
  \begin{minipage}[t]{0.48\linewidth}
    \centering
	\includegraphics[width=8cm]{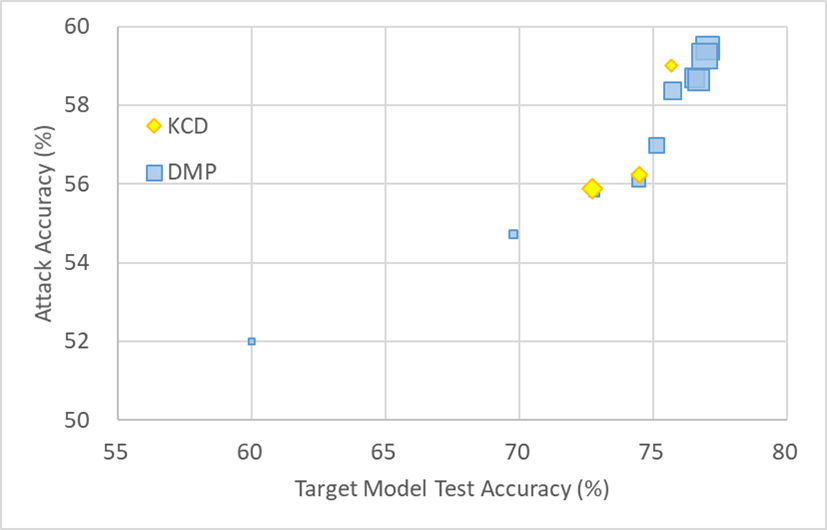}
	\caption{\bf{Privacy-utility trade-off of DMP and KCD for Purchase100.}}\label{fig:num_ref_DMP}
  \end{minipage}
~~
  \begin{minipage}[t]{0.48\linewidth}
    \centering
	\includegraphics[width=8cm]{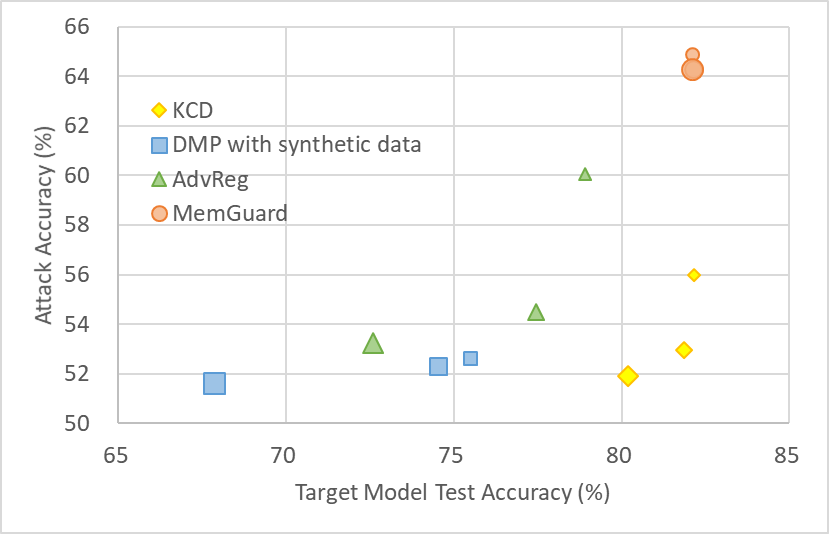}
	\caption{\bf{Privacy-utility trade-offs among defenses for CIFAR10.}
	\rm 
}\label{fig:trade-offs}
\end{minipage}\\
~\\
\begin{center}
\begin{minipage}[t]{0.48\linewidth}
    \centering
    	\includegraphics[width=8cm]{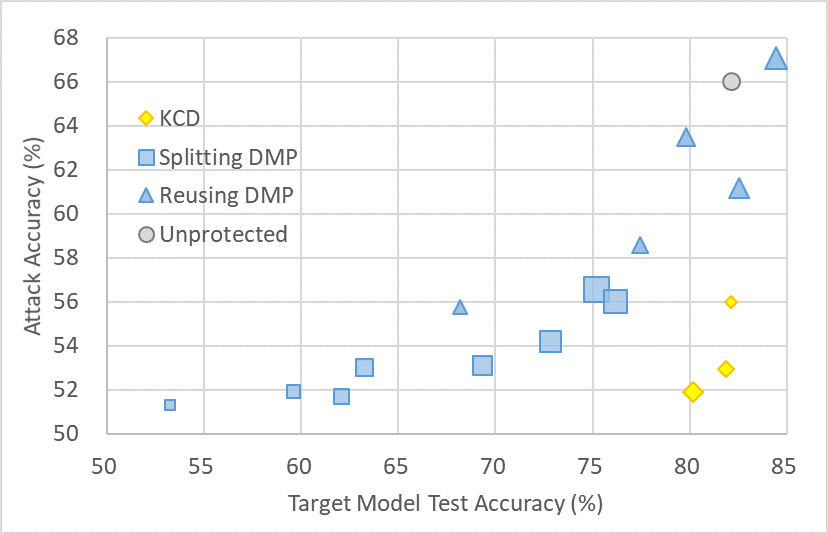}
	\caption{\bf{Privacy-utility trade-off of our KCD, ``splitting'' DMP, ``reusing'' DMP, and unprotected model {for Purchase100}.}
	}\label{fig:naive-DMPvsKCD}
    \end{minipage}\\
\end{center}
{\sf {Explanatory notes on above three figures:}}\small 
\begin{itemize}
	\item {
	The points towards the bottom right are better defenses, i.e., more accurate and more private ones.}
    \item 	{ A larger point means a larger parameter}.
	\item {The vertical axis} {``Attack Accuracy'' means ``BB w/score'' from Table \ref{table:comparison-purchase}}. 
	\item {The privacy-utility trade-off hyperparameters are as follows.}
	\begin{itemize}
        \item {KCD: $\alpha$ of Algorithm \ref{alg:MDMP} ($\alpha=0.6, 0.8, 1.0$ for Purchase100, $\alpha=0.25, 0.5, 1.0$ for CIFAR10)
        \item DMP: the number of pieces of reference data ($n=2000, 4000, \dots, 20000$); this is the best privacy-utility trade-off parameter in our experiment as described in Section \ref{app:tuning} and our $\alpha$ cannot be used directly for DMP as described at the end of Section \ref{subsection:Description}.}
        \item {DMP with synthetic data: the number of pieces of synthetic reference data ($n=12500, 37500, 50000$)}
        \item {AdvReg: the penalty parameter $\lambda=2.623, 3.019, 8.847$}
        \item {MemGuard: the distortion budget $\varepsilon=0.1$, $0.5$, $1.0$} 
        \item {``Splitting'' DMP and ``Reusing'' DMPs: the percentages $\theta$ of training data used to train the student models ($\theta = 7\%$ to $50\%$ for ``splitting'' DMP, $\theta = 20\%$ to $100\%$ for ``reusing'' DMPs). Note that the accuracy of ``reusing DMP'' with $\theta=100\%$ is better than the unprotected model. This kind of increasing accuracy is observed in knowledge distillations \cite{hinton2015distil}.}  
	\end{itemize}
\end{itemize}
\end{figure*}

\section{Experimental Results}
\label{sec:experiment}

\subsection{Tabular Dataset}
\label{subsection:Tabular dataset}

Tables \ref{table:comparison-purchase} and \ref{table:comparison-texas} show the accuracies and MIA attack accuracies of our KCD and known defenses for two tabular datasets, Purchase100 and Texas100.
Here, KCD is compared with the best defenses chosen in Section \ref{subsection:Known_Defenses} from each of the three categories described in the same section.
We stress that one can succeed in an MIA with a probability of 50\% by random guessing. Hence, the baseline of the attack accuracies of these tables is 50\%.

Note that the values for the attack accuracy of MemGuard on these tables are much higher than the values reported in the original paper \cite{jia2019memguard}. This is because the setting we consider, described in Section \ref{subsection:Setting-of-MIA}, is more advantageous for attackers than that of \cite{jia2019memguard}.

Figure \ref{fig:num_ref_DMP} shows the privacy-utility trade-off of KCD and DMP. The results of our experiments for the tabular datasets Purchase100 and Texas100 are summarized as follows.
\begin{enumerate}
    \item Tables \ref{table:comparison-purchase} and \ref{table:comparison-texas} show that KCD was \textit{much better} than the known defenses that also do not use a public reference dataset, AdvReg and MemGuard, in all of the three categories of MIAs, the black-box MIA with confidence score \cite{shokri2017membership, salem2019ml, song2020systematic}, the label-only MIA \cite{choo2020label}, and the white-box MIA \cite{DBLP:conf/sp/NasrSH19}.
    
    For Purchase100, for instance, the testing accuracy of KCD was 11.5\% higher than that of AdvReg and its attack accuracy was 13.3\% smaller than that of MemGuard for ``BB w/score'' attacks.
    \item Surprisingly, Tables \ref{table:comparison-purchase} and \ref{table:comparison-texas} also show that, in both privacy and utility senses and for all of the three categories of MIAs, \textit{KCD is comparable to the state-of-the-art MIA defense, DMP, with public reference data, although KCD does not use public reference data.}
    As mentioned in Section \ref{sec:introduction}, the availability of public data is not guaranteed \cite{shejwalkar2021membership}.
    The above results show that KCD could avoid this problem
    without sacrificing privacy or utility in these experiments.
    \item Figure \ref{fig:num_ref_DMP} shows that, for Purchase100 and for the ``BB w/score'' attack, \textit{the privacy-utility trade-off of KCD was also comparable to that of the state-of-the-art MIA defense requiring public reference data, DMP}. 
    
    We also executed similar experiments for ``label only'' and ``WB.'' They showed that the privacy-utility trade-off of KCD was comparable to that of DMP for these two types of attacks as well. 
\end{enumerate}

\subsection{Image Dataset}

Our experiments for the image dataset CIFAR10 were conducted {in a} similar manner as the above experiment.  We additionally compared KCD with two more defenses. The first one was PATE (with confident-GNMax) \cite{DBLP:conf/iclr/PapernotSMRTE18}. The second one was DMP with synthetic reference data \cite{shejwalkar2021membership} generated using deep convolutional GANs (DCGANs) \cite{dcgan}. Note that {these} two defenses were not used in our experiment with the tabular datasets because they use GANs.

The results of these experiments are summarized as follows.
\begin{enumerate}
    \item Table \ref{table:comparison-CIFAR10} 
    and Figure \ref{fig:trade-offs} show that the privacy-utility trade-off of KCD was \textit{much better} than that of the known defenses without public reference data, AdvReg and MemGuard, and DMP with synthetic reference data for the ``BB w/score'' attack.
    
    We also executed similar experiments for ``label only'' and ``WB.'' They showed that the privacy-utility trade-off of KCD is much better than the known defenses without public reference data as well. 
    \item Table \ref{table:comparison-CIFAR10} also shows that KCD was comparable to DMP and performed much better than PATE in terms of testing accuracy. DMP and PATE were better in terms of privacy, but KCD is better in the sense that it does not require public reference data.
\end{enumerate}

\section{Discussions and Limitations}

\subsection{Discussions}
\label{subsec:Discussions}

\begin{figure}	\includegraphics[width=8cm]{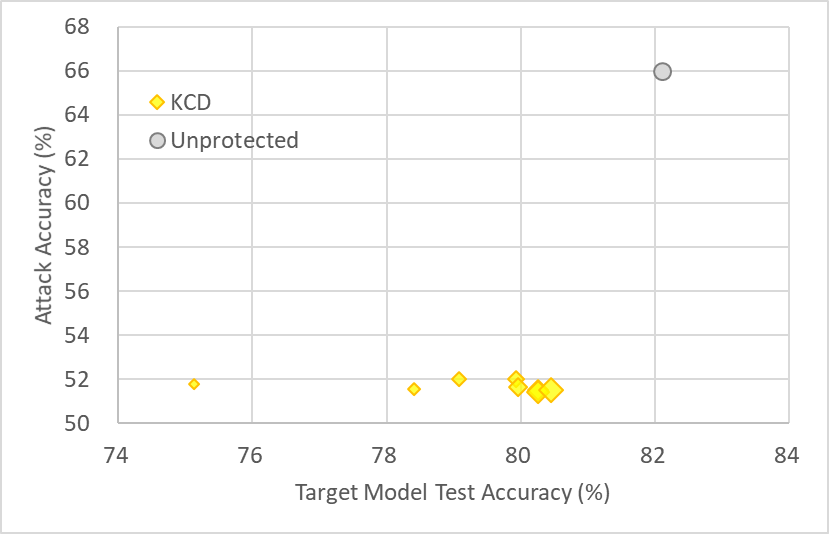}
	\caption{\bf{Effect of number $n$ of teacher models on performance of our KCD for CIFAR10.}
    \rm
    {
    We examined
    $n=2,3,5,7,9,11,13$, and $15$ (a larger point means a larger $n$).
    Note that points indicating $n=11$ and $13$ are too close to distinguish.}
    }\label{fig:num_split_proposed}
\end{figure}

\noindent\textbf{Best number $n$ of teacher models:} Figure~\ref{fig:num_split_proposed} shows

\noindent the performance of our KCD for various teacher models on CIFAR10. Generally, a higher number of teacher models implies better privacy and utility, and they perform the best when $n=15$.

The computational cost of KCD is greater than those of DMP and PATE but less than those of some of the other defense methods, such as AdvReg. A large computational cost may limit  applications for training large models with limited computational resources. However, we believe that the advantage of KCD, ``public reference dataset not necessary,'' makes other applications possible.

We stress that, for inferring, KCD incurs only the same computational cost as an unprotected target model, unlike MemGuard.

\vspace{1em}\noindent\textbf{Comparison with na\"{i}ve ideas:}
To clarify the effect of our ``knowledge cross-distillation'' idea for KCD in terms of privacy and utility, we compared KCD with two na\"{i}ve improvements to DMP to make it ``without reference data.''

The first na\"{i}ve improvement, \textit{``splitting'' DMP}, is as follows. Split the training dataset into two distinct parts; the former and the latter parts contain $(100-\theta)\%$ and $\theta\%$ of training data, respectively. Then, train the teacher model using the former part as a training dataset and train the student model through distillation by using the remaining one as a reference dataset.

The second na\"{i}ve improvement, \textit{``reusing'' DMP}, is as follows. Train a teacher model using all of the training data, take a subset containing $\theta\%$ of training data, and reuse this subset as the reference dataset to train a student model.

Figure \ref{fig:naive-DMPvsKCD} shows that, for the CIFAR10 dataset, the privacy-utility trade-off of our KCD was better than those of these two variants of DMP in our experiments. 
Our KCD contains two ideas,  ``splitting training dataset'' and ``reusing training data for reference data.'' The above result shows that the performance of our KCD is achieved only when both of these ideas are used, and it cannot be achieved with only one of these ideas.

\subsection{Limitations}
\textbf{Duplication in dataset:}
If certain data appear twice in the training dataset, KCD cannot ensure defense against MIAs for such a pair of data. In fact, the defense against MIAs as depicted in Algorithm \ref{alg:MDMP} is ensured because inputs $x\in D_i$ to $F_i$ are not contained in dataset $D\setminus D_i$ used in the training of $F_i$. However, this is not the case when the same data fall into $D_i$ and $D\setminus D_i$, respectively.

Similarly, a training dataset that contains two pieces of data that are not the same but very similar would affect the privacy-utility trade-off of KCD.
Investigating and solving this is for future work.

\vspace{1em}\noindent\textbf{Outlier data, imbalanced dataset:} Long et al.\ \cite{DBLP:journals/corr/abs-1802-04889,DBLP:conf/eurosp/LongWBB0TGC20} showed that an ML model became weaker against MIAs when the target data were outliers or selected carefully by an attacker, even if the ML model was well-generalized.

We selected the target data uniformly at random in our experiments.
Hence, KCD, as well as other known defense methods, may have weak {MIA resistance against carefully selected data}.

Truex et al. \cite{DBLP:conf/tpsisa/TruexLGW019} showed that MIAs against minority classes of imbalanced data were more likely to be successful. Here, imbalanced data means a dataset with skewed class proportions. Minority classes mean the classes that make up a smaller proportion. Hence, KCD, as well as other defense methods, may also have weak protection against MIAs in this case.

\section{Conclusion}

We proposed a new defense against MIAs, \textit{knowledge cross-distillation (KCD)}{, which} does not require any public or synthetic reference data to protect ML models unlike the state-of-the-art defense, DMP.

Our experiments showed that the privacy protection and accuracy of our defense were comparable to those of DMP for the tabular datasets Purchase100 and Texas100, and our defense had a much better privacy-utility trade-off than those of the existing defenses for the CIFAR10 image dataset.

Our defense {is a feasible method} for protecting the privacy of ML models in areas where public reference data are scarce. 
Future work includes ensuring the privacy of duplicated or similar data in a dataset, investigating privacy for outlier and/or imbalanced data, and guaranteeing the privacy of KCD theoretically.

\section*{Acknowledgments}

We would like to thank Reza Shokri,  Tomoyuki Yoshiyama, and Kazuya Kakizaki for useful comments. This research received no specific grant from any funding agency in the public, commercial, or not-for-profit sectors.

\bibliography{acmart}

\begin{thebibliography}{10}

\bibitem{DBLP:conf/sp/Carlini017}
Nicholas Carlini and David~A. Wagner.
\newblock Towards evaluating the robustness of neural networks.
\newblock In {\em {IEEE} Symposium on Security and Privacy}, pages 39--57.
  {IEEE} Computer Society, 2017.

\bibitem{CCPA}
{Bill Text - AB-375 Privacy: personal information: businesses.}
\newblock https://leginfo.legislature.ca.gov/faces/billTextClient.xhtml
  ?bill\_id=201720180AB375, 2018.

\bibitem{chang2019cronus}
Hongyan Chang, Virat Shejwalkar, Reza Shokri, and Amir Houmansadr.
\newblock Cronus: Robust and heterogeneous collaborative learning with
  black-box knowledge transfer.
\newblock {\em arXiv preprint arXiv:1912.11279}, 2019.

\bibitem{chen2020hopskipjumpattack}
Jianbo Chen, Michael~I Jordan, and Martin~J Wainwright.
\newblock Hopskipjumpattack: A query-efficient decision-based attack.
\newblock In {\em 2020 ieee symposium on security and privacy (sp)}, pages
  1277--1294. IEEE, 2020.

\bibitem{choo2020label}
Christopher A~Choquette Choo, Florian Tramer, Nicholas Carlini, and Nicolas
  Papernot.
\newblock Label-only membership inference attacks.
\newblock {\em arXiv preprint arXiv:2007.14321}, 2020.

\bibitem{dlsurvey}
Shaveta Dargan, Munish Kumar, Maruthi~Rohit Ayyagari, and Gulshan Kumar.
\newblock A survey of deep learning and its applications: A new paradigm to
  machine learning.
\newblock {\em Archives of Computational Methods in Engineering}, pages 1--22,
  2019.

\bibitem{10.1007/11787006_1}
Cynthia Dwork.
\newblock Differential privacy.
\newblock In Michele Bugliesi, Bart Preneel, Vladimiro Sassone, and Ingo
  Wegener, editors, {\em Automata, Languages and Programming}, pages 1--12,
  Berlin, Heidelberg, 2006. Springer Berlin Heidelberg.

\bibitem{DBLP:conf/eurocrypt/DworkKMMN06}
Cynthia Dwork, Krishnaram Kenthapadi, Frank McSherry, Ilya Mironov, and Moni
  Naor.
\newblock Our data, ourselves: Privacy via distributed noise generation.
\newblock In Serge Vaudenay, editor, {\em Advances in Cryptology - {EUROCRYPT}
  2006, 25th Annual International Conference on the Theory and Applications of
  Cryptographic Techniques, St. Petersburg, Russia, May 28 - June 1, 2006,
  Proceedings}, volume 4004 of {\em Lecture Notes in Computer Science}, pages
  486--503. Springer, 2006.

\bibitem{EU-Guideline}
Ethics {G}uidelines for {T}rustworthy {AI}.
\newblock
  https://digital-strategy.ec.europa.eu/en/library/ethics-guidelines-trustworthy-ai,
  2019.

\bibitem{GDPR}
{REGULATION (EU) 2016/ 679 OF THE EUROPEAN PARLIAMENT AND OF THE COUNCIL - of
  27 April 2016 - on the protection of natural persons with regard to the
  processing of personal data and on the free movement of such data, and
  repealing Directive 95/ 46/ EC (General Data Protection Regulation)}.
\newblock
  https://eur-lex.europa.eu/legal-content/EN/TXT/PDF/?uri=CELEX:32016R0679,
  2016.

\bibitem{HayesMDC19LOGAN}
Jamie Hayes, Luca Melis, George Danezis, and Emiliano~De Cristofaro.
\newblock {LOGAN:} membership inference attacks against generative models.
\newblock {\em Proc. Priv. Enhancing Technol.}, 2019(1):133--152, 2019.

\bibitem{hinton2015distil}
Geoffrey Hinton, Oriol Vinyals, and Jeffrey Dean.
\newblock Distilling the knowledge in a neural network.
\newblock In {\em NIPS Deep Learning and Representation Learning Workshop},
  2015.

\bibitem{HIPAA}
{HIPAA}.
\newblock
  https://www.govinfo.gov/content/pkg/PLAW-104publ191/pdf/PLAW-104publ191.pdf,
  1996.

\bibitem{homer2008resolving}
Nils Homer, Szabolcs Szelinger, Margot Redman, David Duggan, Waibhav Tembe,
  Jill Muehling, John~V. Pearson, Dietrich~A. Stephan, Stanley~F. Nelson, and
  David~W. Craig.
\newblock Resolving individuals contributing trace amounts of {DNA} to highly
  complex mixtures using high-density {SNP} genotyping microarrays.
\newblock {\em PLoS Genet}, 4(8):e1000167, 08 2008.

\bibitem{DBLP:journals/corr/abs-2103-07853}
Hongsheng Hu, Zoran Salcic, Gillian Dobbie, and Xuyun Zhang.
\newblock Membership inference attacks on machine learning: A survey.
\newblock {\em CoRR}, abs/2103.07853, 2021.

\bibitem{DBLP:conf/ndss/HuiYYBGC21}
Bo~Hui, Yuchen Yang, Haolin Yuan, Philippe Burlina, Neil~Zhenqiang Gong, and
  Yinzhi Cao.
\newblock Practical blind membership inference attack via differential
  comparisons.
\newblock In {\em 28th Annual Network and Distributed System Security
  Symposium, {NDSS} 2021, virtually, February 21-25, 2021}. The Internet
  Society, 2021.

\bibitem{jia2019memguard}
Jinyuan Jia, Ahmed Salem, Michael Backes, Yang Zhang, and Neil~Zhenqiang Gong.
\newblock Memguard: Defending against black-box membership inference attacks
  via adversarial examples.
\newblock In {\em Proceedings of the 2019 ACM SIGSAC Conference on Computer and
  Communications Security}, pages 259--274, 2019.

\bibitem{Krizhevsky09}
A.~Krizhevsky and G.~Hinton.
\newblock Learning multiple layers of features from tiny images.
\newblock {\em Master's thesis, Department of Computer Science, University of
  Toronto}, 2009.

\bibitem{DBLP:conf/uss/LeinoF20}
Klas Leino and Matt Fredrikson.
\newblock Stolen memories: Leveraging model memorization for calibrated
  white-box membership inference.
\newblock In Srdjan Capkun and Franziska Roesner, editors, {\em 29th {USENIX}
  Security Symposium, {USENIX} Security 2020, August 12-14, 2020}, pages
  1605--1622. {USENIX} Association, 2020.

\bibitem{DBLP:journals/corr/abs-2007-15528}
Zheng Li and Yang Zhang.
\newblock Label-leaks: Membership inference attack with label.
\newblock {\em CoRR}, abs/2007.15528, 2020.

\bibitem{DBLP:journals/corr/abs-1802-04889}
Yunhui Long, Vincent Bindschaedler, Lei Wang, Diyue Bu, Xiaofeng Wang, Haixu
  Tang, Carl~A. Gunter, and Kai Chen.
\newblock Understanding membership inferences on well-generalized learning
  models.
\newblock {\em CoRR}, abs/1802.04889, 2018.

\bibitem{DBLP:conf/eurosp/LongWBB0TGC20}
Yunhui Long, Lei Wang, Diyue Bu, Vincent Bindschaedler, XiaoFeng Wang, Haixu
  Tang, Carl~A. Gunter, and Kai Chen.
\newblock A pragmatic approach to membership inferences on machine learning
  models.
\newblock In {\em {IEEE} European Symposium on Security and Privacy, EuroS{\&}P
  2020, Genoa, Italy, September 7-11, 2020}, pages 521--534. {IEEE}, 2020.

\bibitem{nasr2018machine}
Milad Nasr, Reza Shokri, and Amir Houmansadr.
\newblock Machine learning with membership privacy using adversarial
  regularization.
\newblock In {\em Proceedings of the 2018 ACM SIGSAC Conference on Computer and
  Communications Security}, pages 634--646, 2018.

\bibitem{DBLP:conf/sp/NasrSH19}
Milad Nasr, Reza Shokri, and Amir Houmansadr.
\newblock Comprehensive privacy analysis of deep learning: Passive and active
  white-box inference attacks against centralized and federated learning.
\newblock In {\em {IEEE} Symposium on Security and Privacy}, pages 739--753.
  {IEEE}, 2019.

\bibitem{Optuna}
Optuna.
\newblock https://optuna.org/.

\bibitem{DBLP:conf/iclr/PapernotAEGT17}
Nicolas Papernot, Mart{\'{\i}}n Abadi, {\'{U}}lfar Erlingsson, Ian~J.
  Goodfellow, and Kunal Talwar.
\newblock Semi-supervised knowledge transfer for deep learning from private
  training data.
\newblock In {\em {ICLR}}. OpenReview.net, 2017.

\bibitem{DBLP:conf/iclr/PapernotSMRTE18}
Nicolas Papernot, Shuang Song, Ilya Mironov, Ananth Raghunathan, Kunal Talwar,
  and {\'{U}}lfar Erlingsson.
\newblock Scalable private learning with {PATE}.
\newblock In {\em 6th International Conference on Learning Representations,
  {ICLR} 2018, Vancouver, BC, Canada, April 30 - May 3, 2018, Conference Track
  Proceedings}. OpenReview.net, 2018.

\bibitem{Purchase100}
{Kaggle’s Acquire Valued Shoppers Challenge}.
\newblock https://www.kaggle.com/c/acquire-valued-shoppers-challenge, 2013.

\bibitem{dcgan}
Alec Radford, Luke Metz, and Soumith Chintala.
\newblock Unsupervised representation learning with deep convolutional
  generative adversarial networks, 2016.

\bibitem{sablayrolles2019white}
Alexandre Sablayrolles, Matthijs Douze, Yann Ollivier, Cordelia Schmid, and
  Herv{\'e} J{\'e}gou.
\newblock White-box vs black-box: Bayes optimal strategies for membership
  inference.
\newblock In {\em ICML 2019-36th International Conference on Machine Learning},
  volume~97, pages 5558--5567, 2019.

\bibitem{salem2019ml}
Ahmed Salem, Yang Zhang, Mathias Humbert, Mario Fritz, and Michael Backes.
\newblock Ml-leaks: Model and data independent membership inference attacks and
  defenses on machine learning models.
\newblock In {\em Network and Distributed Systems Security Symposium 2019}.
  Internet Society, 2019.

\bibitem{NIPS2016_8a3363ab}
Tim Salimans, Ian Goodfellow, Wojciech Zaremba, Vicki Cheung, Alec Radford,
  Xi~Chen, and Xi~Chen.
\newblock Improved techniques for training gans.
\newblock In D.~Lee, M.~Sugiyama, U.~Luxburg, I.~Guyon, and R.~Garnett,
  editors, {\em Advances in Neural Information Processing Systems}, volume~29.
  Curran Associates, Inc., 2016.

\bibitem{shejwalkar2021membership}
Virat Shejwalkar and Amir Houmansadr.
\newblock Membership privacy for machine learning models through knowledge
  transfer.
\newblock In {\em Thirty-Fifth {AAAI} Conference on Artificial Intelligence,
  {AAAI} 2021, Thirty-Third Conference on Innovative Applications of Artificial
  Intelligence, {IAAI} 2021, The Eleventh Symposium on Educational Advances in
  Artificial Intelligence, {EAAI} 2021, Virtual Event, February 2-9, 2021},
  pages 9549--9557. {AAAI} Press, 2021.

\bibitem{shokri2017membership}
Reza Shokri, Marco Stronati, Congzheng Song, and Vitaly Shmatikov.
\newblock Membership inference attacks against machine learning models.
\newblock In {\em 2017 IEEE Symposium on Security and Privacy (SP)}, pages
  3--18. IEEE, 2017.

\bibitem{song2020systematic}
Liwei Song and Prateek Mittal.
\newblock Systematic evaluation of privacy risks of machine learning models.
\newblock {\em arXiv preprint arXiv:2003.10595}, 2020.
\newblock (Accepted in USENIX Security 2021.).

\bibitem{DBLP:conf/ccs/SongSM19}
Liwei Song, Reza Shokri, and Prateek Mittal.
\newblock Privacy risks of securing machine learning models against adversarial
  examples.
\newblock In Lorenzo Cavallaro, Johannes Kinder, XiaoFeng Wang, and Jonathan
  Katz, editors, {\em Proceedings of the 2019 {ACM} {SIGSAC} Conference on
  Computer and Communications Security, {CCS} 2019, London, UK, November 11-15,
  2019}, pages 241--257. {ACM}, 2019.

\bibitem{DBLP:journals/jmlr/SrivastavaHKSS14}
Nitish Srivastava, Geoffrey~E. Hinton, Alex Krizhevsky, Ilya Sutskever, and
  Ruslan Salakhutdinov.
\newblock Dropout: a simple way to prevent neural networks from overfitting.
\newblock {\em J. Mach. Learn. Res.}, 15(1):1929--1958, 2014.

\bibitem{DBLP:journals/corr/SzegedyZSBEGF13}
Christian Szegedy, Wojciech Zaremba, Ilya Sutskever, Joan Bruna, Dumitru Erhan,
  Ian~J. Goodfellow, and Rob Fergus.
\newblock Intriguing properties of neural networks.
\newblock In Yoshua Bengio and Yann LeCun, editors, {\em 2nd International
  Conference on Learning Representations, {ICLR} 2014, Banff, AB, Canada, April
  14-16, 2014, Conference Track Proceedings}, 2014.

\bibitem{tang2021mitigating}
Xinyu Tang, Saeed Mahloujifar, Liwei Song, Virat Shejwalkar, Milad Nasr, Amir
  Houmansadr, and Prateek Mittal.
\newblock Mitigating membership inference attacks by self-distillation through
  a novel ensemble architecture.
\newblock {\em arXiv preprint arXiv:2110.08324}, 2021.

\bibitem{Texas100}
{Hospital Discharge Data Public Use Data File}.
\newblock https://www.dshs.texas.gov/THCIC/Hospitals/\\Download.shtm.

\bibitem{DBLP:conf/tpsisa/TruexLGW019}
Stacey Truex, Ling Liu, Mehmet~Emre Gursoy, Wenqi Wei, and Lei Yu.
\newblock Effects of differential privacy and data skewness on membership
  inference vulnerability.
\newblock In {\em First {IEEE} International Conference on Trust, Privacy and
  Security in Intelligent Systems and Applications, {TPS-ISA} 2019, Los
  Angeles, CA, USA, December 12-14, 2019}, pages 82--91. {IEEE}, 2019.

\bibitem{8634878}
Stacey Truex, Ling Liu, Mehmet~Emre Gursoy, Lei Yu, and Wenqi Wei.
\newblock Demystifying membership inference attacks in machine learning as a
  service.
\newblock {\em IEEE Transactions on Services Computing}, pages 1--1, 2019.

\bibitem{veale2018algorithms}
Michael Veale, Reuben Binns, and Lilian Edwards.
\newblock Algorithms that remember: model inversion attacks and data protection
  law.
\newblock {\em Philosophical Transactions of the Royal Society A: Mathematical,
  Physical and Engineering Sciences}, 376(2133):20180083, 2018.

\bibitem{yeom2018privacy}
Samuel Yeom, Irene Giacomelli, Matt Fredrikson, and Somesh Jha.
\newblock Privacy risk in machine learning: Analyzing the connection to
  overfitting.
\newblock In {\em 2018 IEEE 31st Computer Security Foundations Symposium
  (CSF)}, pages 268--282. IEEE, 2018.

\end{thebibliography}
\bibliographystyle{plain}

\onecolumn
\appendix
\section{Appendix}
\label{section:Missing_Details_of_Experimental_Results}
\begin{table}[h]
 \begin{tabular}{l|l|l||l|l|l|l|l|l}
 \hline
 \multicolumn{1}{|c|}{ }& \rm Category & \rm Defense&\rm \!\!Leaks~1~\cite{salem2019ml}\!\! & \rm \!\!Top~1~\cite{salem2019ml}\!\!& \rm \!\!Corr.~\cite{song2020systematic}\!\!& \rm \!\!Conf.~\cite{song2020systematic}\!\!& \rm \!\!Entr.~\cite{song2020systematic}\!\!& \multicolumn{1}{c|}{\rm \!\!m-Entr.~\cite{song2020systematic}\!\!}\\\hline\hline
 \multicolumn{1}{|c|}{\multirow{3}{*}{\rm ${}^\nexists$Public Ref.}}& \rm Reg-based& \rm AdvReg \cite{nasr2018machine}& \rm 57.0\% & \rm 56.8\% & \rm 58.9\% & \rm 59.9\% & \rm 55.3\% & \multicolumn{1}{c|}{\rm 59.7\% }\\
 \cline{2-9}
 \multicolumn{1}{|c|}{ }& \rm AX-based& \rm MemGuard \cite{jia2019memguard}& \rm 66.6\% & \rm 71.9\% & \rm 61.3\% & \rm 72.1\% & \rm 70.1\% & \multicolumn{1}{c|}{\rm 72.1\% } \\ 
 \cline{2-9}
 \multicolumn{1}{|c|}{ }& \rm KT-based& \bf KCD& \rm 54.9\% & \rm 55.0\% & \rm 58.8\% & \rm 57.0\% & \rm 53.7\% & \multicolumn{1}{c|}{\rm 57.3\% }\\
 \hline
 \multicolumn{1}{|c|}{\rm ${}^\exists$Public Ref.}& \rm KT-based& \rm DMP~\cite{shejwalkar2021membership}\quad\quad\quad\quad\quad\quad& \rm 53.9\% & \rm 53.9\% & \rm 57.1\% & \rm 55.8\% & \rm 52.8\% & \multicolumn{1}{c|}{\rm 55.7\% }\\
 \hline\hline
 \multicolumn{3}{|c||}{\rm Unprotected} & \rm 72.8\% & \rm 72.0\% & \rm 61.3\% & \rm 73.6\% & \rm 71.2\% & \multicolumn{1}{c|}{\rm 73.7\% }\\
 \hline
 \end{tabular}
 \caption{\centering BB attacks with confidence scores on Purchase100}
 \label{table:BB-purchase100}

 \begin{tabular}{l|l|l||l|l|l|l|l|l}
 \hline
 \multicolumn{1}{|c|}{ }& \rm Category & \rm Defense&\rm \!\!Leaks~1~\cite{salem2019ml}\!\! & \rm \!\!Top~1~\cite{salem2019ml}\!\!& \rm \!\!Corr.~\cite{song2020systematic}\!\!& \rm \!\!Conf.~\cite{song2020systematic}\!\!& \rm \!\!Entr.~\cite{song2020systematic}\!\!& \multicolumn{1}{c|}{\rm \!\!m-Entr.~\cite{song2020systematic}\!\!}\\\hline\hline
 \multicolumn{1}{|c|}{\multirow{3}{*}{\rm ${}^\nexists$Public Ref.}}& \rm Reg-based& \rm AdvReg \cite{nasr2018machine}& \rm 52.5\% & \rm 52.1\% & \rm 56.7\% & \rm 58.8\% & \rm 53.2\% & \multicolumn{1}{c|}{\rm 59.5\% }\\
 \cline{2-9}
 \multicolumn{1}{|c|}{ }& \rm AX-based& \rm MemGuard \cite{jia2019memguard}& \rm 57.7\% & \rm 58.0\% & \rm 68.6\% & \rm 68.2\% & \rm 57.7\% & \multicolumn{1}{c|}{\rm 68.2\% } \\ 
 \cline{2-9}
 \multicolumn{1}{|c|}{ }& \rm KT-based& \bf KCD& \rm 54.8\% & \rm 54.9\% & \rm 53.1\% & \rm 56.2\% & \rm 54.8\% & \multicolumn{1}{c|}{\rm 55.4\% }\\
 \hline
 \multicolumn{1}{|c|}{\rm ${}^\exists$Public Ref.}& \rm KT-based& \rm DMP \cite{shejwalkar2021membership}\quad\quad\quad\quad\quad\quad& \rm 51.2\% & \rm 51.5\% & \rm 56.1\% & \rm 56.3\% & \rm 51.0\% & \multicolumn{1}{c|}{\rm 56.1\% }\\
 \hline\hline
 \multicolumn{3}{|c||}{\rm Unprotected} & \rm 58.8\% & \rm 58.6\% & \rm 68.6\% & \rm 69.7\% & \rm 59.4\% & \multicolumn{1}{c|}{\rm 69.9\% }\\
 \hline
 \end{tabular}
\caption{\centering BB attacks with confidence scores on Texas100}
 \label{table:BB-texas100}

 \begin{tabular}{l|l|l||l|l|l|l|l|l}
 \hline
 \multicolumn{1}{|c|}{ }& \rm Category & \rm Defense&\rm \!\!Leaks~1~\cite{salem2019ml}\!\! & \rm \!\!Top~1~\cite{salem2019ml}\!\!& \rm \!\!Corr.~\cite{song2020systematic}\!\!& \rm \!\!Conf.~\cite{song2020systematic}\!\!& \rm \!\!Entr.~\cite{song2020systematic}\!\!& \multicolumn{1}{c|}{\rm \!\!m-Entr.~\cite{song2020systematic}\!\!}\\
\hline\hline
 \multicolumn{1}{|c|}{\multirow{4}{*}{\rm ${}^\nexists$Public Ref.}}& \rm Reg-based& \rm AdvReg \cite{nasr2018machine}& \rm 53.1\% & \rm 52.7\% & \rm 54.6\% & \rm 54.6\% & \rm 51.9\% & \multicolumn{1}{c|}{\rm 54.6\% }\\
 \cline{2-9}
 \multicolumn{1}{|c|}{ }& \rm AX-based& \rm MemGuard \cite{jia2019memguard}& \rm 63.0\% & \rm 63.4\% & \rm 58.6\% & \rm 64.3\% & \rm 63.1\% & \multicolumn{1}{c|}{\rm 64.3\% } \\ 
 \cline{2-9}
 \multicolumn{1}{|c|}{ }& \multicolumn{1}{c|}{\multirow{2}{*}{\rm KT-based}}& \rm DMP \cite{shejwalkar2021membership}(synth. ref.)& \rm 51.0\% & \rm 51.2\% & \rm 52.5\% & \rm 51.8\% & \rm 50.3\% & \multicolumn{1}{c|}{\rm 52.0\% }\\
 \cline{3-9}
 \multicolumn{1}{|c|}{ }& \rm & \bf KCD& \rm 52.1\% & \rm 52.2\% & \rm 55.6\% & \rm 55.3\% & \rm 51.3\% & \multicolumn{1}{c|}{\rm 55.8\% }\\
 \hline
 \multicolumn{1}{|c|}{\multirow{2}{*}{\rm ${}^\exists$Public Ref.}}& \multicolumn{1}{c|}{\multirow{2}{*}{\rm KT-based}}& \rm DMP \cite{shejwalkar2021membership}& \rm 50.8\% & \rm 50.7\% & \rm 50.7\% & \rm 50.4\% & \rm 51.1\% & \multicolumn{1}{c|}{\rm 50.2\%}\\
 \cline{3-9}
 \multicolumn{1}{|c|}{ }& & \rm PATE \cite{DBLP:conf/iclr/PapernotSMRTE18}& \rm 50.0\% &\rm 49.8\% &\rm 50.5\% &\rm 50.4\% &\rm 50.0\% & \multicolumn{1}{c|}{\rm 51.2\%}\\
 \hline\hline
 \multicolumn{3}{|c||}{\rm Unprotected}& \rm 64.2\% & \rm 63.8\% & \rm 58.6\% & \rm 65.6\% & \rm 63.9\% & \multicolumn{1}{c|}{\rm 65.9\% }\\
 \hline
 \end{tabular}
\caption{\centering BB attacks with confidence scores on CIFAR10}
 \label{table:BB-cifar10}

The above tables show the attack accuracies of each black-box MIA with confidence scores. ``Leaks 1'' means ML Leaks Adversary 1 \cite{salem2019ml}. ``Top 1,'' ``Corr.,'' ``Conf.,'' ``Entr.,'' ``m-Entr.,'' mean five metric-based attacks \cite{salem2019ml,song2020systematic}, \textit{Top 1}, \textit{correctness}, \textit{confidence}, \textit{entropy}, and \textit{m-entropy attacks}, respectively.
\end{table}

\begin{figure}
\centering
 \begin{minipage}[t]{0.50\linewidth}
\includegraphics[width=8cm]{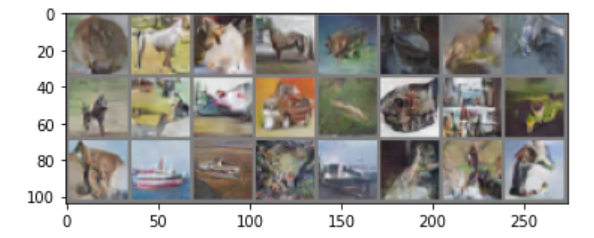}
\caption{Images generated by (unconditional) DCGAN}\label{fig:generated_cifar10}
\end{minipage}
\end{figure}

\makeatletter
\renewcommand{\ALG@name}{Algorithm}
\newenvironment{breakablealgorithm}
{
\begin{center}
\refstepcounter{algorithm}
\hrule height.8pt depth0pt \kern2pt
\renewcommand{\caption}[2][\relax]{
{\raggedright\textbf{\ALG@name~\thealgorithm} ##2\par}%
\ifx\relax##1\relax 
\addcontentsline{loa}{algorithm}{\protect\numberline{\thealgorithm}##2}%
\else 
\addcontentsline{loa}{algorithm}{\protect\numberline{\thealgorithm}##1}%
\fi
\kern2pt\hrule\kern2pt
}
}{
\kern2pt\hrule\relax
\end{center}
}
\makeatother

\end{document}